# Comment: Gibbs Sampling, Exponential Families, and Orthogonal Polynomials

**Galin L. Jones and Alicia A. Johnson**

It is our pleasure to congratulate the authors (hereafter DKSC) on an interesting paper that was a delight to read. While DKSC provide a remarkable collection of connections between different representations of the Markov chains in their paper, we will focus on the "running time analysis" portion. This is a familiar problem to statisticians; given a target population, how can we obtain a representative sample? In the context of Markov chain Monte Carlo (MCMC) the problem can be stated as follows. Let $\Phi = \{X_0, X_1, X_2, \ldots\}$ be an irreducible aperiodic Markov chain with invariant probability distribution $\pi$ having support $\mathcal{X}$ and let $P^n$ denote the distribution of $X_n \mid X_0$ for $n \geq 1$, that is, $P^n(x, A) = \Pr(X_n \in A \mid X_0 = x)$. Then, given $\omega > 0$, can we find a positive integer $n^*$ such that

$$\|P^{n^*}(x, \cdot) - \pi(\cdot)\| \leq \omega \qquad (1)$$

where $\|\cdot\|$ is the total variation norm? If we can find such an $n^*$, then, since $\|P^n - \pi\|$ is nonincreasing in $n$, every draw past $n^*$ will also be within $\omega$ of $\pi$, thus providing a representative sample if we keep only the draws after $n^*$. There is an enormous amount of research (too much to list here!) on this problem for a wide variety of Markov chains. Unfortunately, there is apparently little that can be said generally about this problem so that we are forced to analyze each Markov chain individually or at most within a limited class of models or situations. This is somewhat reflected in the current paper since, as noted by DKSC, the techniques introduced here do not apply to even all of the exponential families (with a conjugate prior) in the paper. However, DKSC derive some impressive results that would seem difficult to improve upon. In the rest of this discussion we will review some of their findings and compare them to results possible via the so-called (by DKSC) "Harris recurrence techniques."

## 1. FINITE SAMPLE VS. ASYMPTOTIC

Perhaps the most common use of MCMC by statisticians is to estimate an expectation with respect to $\pi$. More specifically, suppose $g : \mathcal{X} \to \mathbb{R}$ and our goal is to calculate

$$E_\pi g = \int_{\mathcal{X}} g(x) \pi(dx).$$

In typical MCMC settings, this quantity is analytically intractable and we estimate it with a sample average

$$\bar{g}_{n,B} = \frac{1}{n} \sum_{i=B}^{n+B-1} g(X_i)$$

over the observed path of the Markov chain. Here $B$ denotes the burn-in. If we can find $n^*$ satisfying (1) it would be natural to set $B = n^*$ thereby reducing the inherent bias in $\bar{g}_{n,B}$ but possibly increasing its variance compared to using all $n+B$ draws. In any case, $\bar{g}_{n,B}$ is a useful estimator since as long as $E_\pi |g| < \infty$, we have a strong law: $\bar{g}_{n,B} \to E_\pi g$ with probability 1 as $n \to \infty$.

Of course, no matter what the simulation length (i.e. $n+B$) there will be an unknown Monte Carlo error in our estimate, namely $\bar{g}_{n,B} - E_\pi g$. When it holds, a Markov chain central limit theorem (CLT) provides an approximate sampling distribution of the Monte Carlo error as well as a basis for finding the corresponding Monte Carlo standard error; see Flegal, Haran and Jones (2008), Jones et al. (2006) and Jones and Hobert (2001).

Thus, we see that there are two questions we want to answer in order that we handle the output of the sampler in a sensible fashion. Specifically, what is $n^*$ and when does a Markov chain CLT hold? At first glance these two properties are seemingly unrelated. That is, finding $n^*$ is about a finite-sample property of the Markov chain while the existence of a CLT is asymptotic. In fact, they are not so different. A key sufficient condition for the existence of a Markov chain CLT is that the Markov chain is geometrically ergodic. That is, there exists $M : \mathcal{X} \to \mathbb{R}$ and a constant $t \in (0, 1)$ such that

$$(2) \qquad \|P^n(x, \cdot) - \pi(\cdot)\| \leq M(x) t^n.$$

*Galin L. Jones is Associate Professor, School of Statistics, University of Minnesota, Minneapolis, Minnesota, USA e-mail: galin@stat.umn.edu. Alicia A. Johnson is Graduate Student, School of Statistics, University of Minnesota, Minneapolis, Minnesota, USA e-mail: ajohnson@stat.umn.edu.*

(See Jones (2004), and Roberts and Rosenthal (2004), for much more on Markov chain CLTs.) Note that as long as the initial value of the simulation is not chosen too poorly, that is $M(x)$ is not too large, geometric ergodicity ensures rapid convergence. Also, if we could find $M$ and $t$ satisfying (2), then a CLT would hold as long as $E_\pi |g|^{2+\delta} < \infty$ for some $\delta > 0$ and we could easily use (2) to find $n^*$. Unfortunately, $M$ and $t$ are rarely available in practically relevant settings, so that the best we can hope to do is find bounds for them. In the next section we consider a constructive method for establishing (2), that is the existence of $M$ and $t$.

## 2. DRIFT AND MINORIZATION

A *drift condition* holds if there exists some function $V : \mathcal{X} \to [0, \infty)$ and constants $0 < \gamma < 1$ and $L < \infty$ such that

$$(3) \qquad \mathrm{E}[V(X_{i+1}) \mid X_i = x] \leq \gamma V(x) + L$$

for all $x \in \mathcal{X}$.

The set $C \subseteq \mathcal{X}$ is *small* if there exists a probability measure $Q$ on $\mathcal{X}$ and some $\varepsilon > 0$ for which the following *minorization condition* holds:

$$(4) \qquad P(x, A) \geq \varepsilon Q(A)$$

for all $x \in C$ and $A \in \mathcal{B}(\mathcal{X})$.

If a drift condition holds and the set $C = \{x : V(x) \leq w\}$ for $w > 2L/(1 - \gamma)$ is small, then the Markov chain is geometrically ergodic. Rosenthal (1995) showed that in this case, the drift and minorization conditions can also be used to find a value of $n^*$ satisfying (1); see also Baxendale (2005), Hobert and Robert (2004) and Roberts and Tweedie (1999). While Rosenthal's theorem often results in conservative values for $n^*$, they can still be useful. We will illustrate this in a simple example below;

see Jones and Hobert (2004) for a practically relevant example.

There are at least two other interesting implications of drift and minorization: Kendall (2004) showed that the existence of drift and minorization imply the existence of a perfect sampling algorithm and Latuszynski (2008) recently has shown that drift and minorization can be used to find a simulation length that will guarantee that $\bar{g}_{n,B}$ is within a user-specified distance of $E_\pi g$ with a user-specified probability.

## 3. DRIFT FOR EXPONENTIAL FAMILIES

Following Section 2.3.1 of DKSC, we will assume that the distribution of $X \mid \theta$ is from one of the six exponential families and use the conjugate prior for $\theta$. Let $m(\cdot)$ be the marginal density of $X$. Also, let $\Phi = \{X_0, X_1, X_2, \ldots\}$ denote the $x$ chain for these families with one-step and $l$-step transition densities $k_x(\cdot)$ and $k_x^l(\cdot)$, respectively. DKSC construct sharp bounds on the total variation distance of the chain to stationarity, $\|k_x^l - m\|$. Our goal is to construct a drift and associated minorization condition for the $x$ chain. As discussed above this will allow us to conclude the Markov chains are geometrically ergodic and compare the value of $n^*$ given by Rosenthal's theorem to that obtained by DKSC.

As in DKSC, the conditional expectations of $X$ and $\theta$ have a special form when $\theta$ is assigned a conjugate prior, specifically, $\mathrm{E}(X^k \mid \theta)$ and $\mathrm{E}(\theta^k \mid X)$ "are polynomials of degree $k$ in $\theta$ and $X$ respectively." That is, we can define constants so that

$$(5) \qquad \begin{aligned} \mathrm{E}(X \mid \theta) &= a\theta + b, \\ \mathrm{E}(\theta \mid X) &= fX + g, \\ \mathrm{E}(X^2 \mid \theta) &= c\theta^2 + d\theta + e, \\ \mathrm{E}(\theta^2 \mid X) &= hX^2 + jX + k. \end{aligned}$$

PROPOSITION 1. *Using the notation defined in (5), assume $ch < 1$ and define*

$$u = \frac{df + cj}{2(af - ch)}.$$

*Fix $\gamma \in [ch, 1)$. Then the following drift condition is satisfied for the $x$-chain making transition $x \to Y$:*

$$E[V(Y) \mid x] \leq \gamma V(x) + L$$

*where $V(Y) = (Y - u)^2$ and $L = ck + gd + e + u^2(1 - ch) - 2u(ga + b)$.*

PROOF. First, notice that $E[V(Y) \mid x] = E[(Y - u)^2 \mid x]$ where

$$E[(Y-u)^2 \mid x]$$
$$= E[E[(Y-u)^2 \mid \theta] \mid x]$$
$$= E[\text{Var}[(Y-u) \mid \theta] + \{E[(Y-u) \mid \theta]\}^2 \mid x]$$
$$= E[E(Y^2 \mid \theta) - [E(Y \mid \theta)]^2 + [E(Y \mid \theta) - u]^2 \mid x]$$
$$= E[E(Y^2 \mid \theta) - 2uE(Y \mid \theta) + u^2 \mid x].$$

Combining this and (5) with some algebra gives

$$E[E(Y^2 \mid \theta) - 2uE(Y \mid \theta) + u^2 \mid x]$$
$$= E[c\theta^2 + \theta(d - 2ua) \mid x] + [e - 2ub + u^2]$$
$$= ch(x - u)^2 + L.$$

Therefore,

$$E[V(Y) \mid x] = chV(x) + L \le \gamma V(x) + L$$

and the result holds. □

REMARK 1. Proposition 1 holds for each of the Beta/Binomial, Poisson/Gamma and Gaussian families. However, restrictions must be placed on the hyperparameters of the remaining three exponential families to ensure $ch < 1$.

REMARK 2. We are making no claim that the above drift condition is optimal for all (or even any) of the exponential families considered in DKSC. In fact, it is easy to cook up many more functions for which (3) is easily verified, especially if each family is considered individually. We prefer drift functions satisfying (3) that lead to larger values of $\varepsilon$ in (4) for $C = \{x : V(x) \le w\}$.

It is possible to establish a minorization condition for the setting so far described. However, we have found that exploiting the structure of a particular example often leads to a larger value of $\varepsilon$. As an illustration we will focus on the Gaussian setting in the next section.

## 4. MINORIZATION IN THE GAUSSIAN EXAMPLE

Consider the Gaussian model where $\nu = 0$ and $\sigma^2 = \tau^2 = 1/4$, that is,

$$X \mid \theta \sim N(\theta, 1/4) \quad \text{and} \quad \theta \sim N(0, 1/4).$$

From here, it is straightforward to show that

$$X \sim N(0, 1/2) \quad \text{and} \quad \theta \mid X \sim N(X/2, 1/8).$$

A similar Gibbs sampler is analyzed in example 1 of Rosenthal (1995). In this case, $a = c = 1$, $b = d = g = j = 0$, $e = 1/4$, $f = 1/2$, $h = 1/4$ and $k = 1/8$ since

$$E(X \mid \theta) = \theta,$$
$$E(X^2 \mid \theta) = \theta^2 + \tfrac{1}{4},$$
$$E(\theta \mid X) = \tfrac{1}{2}X,$$
$$E(\theta^2 \mid X) = \tfrac{1}{4}X^2 + \tfrac{1}{8}.$$

Notice that $ch = 1/4$, $u = 0$, and $L = 3/8$ where $u$ and $L$ are as defined by Proposition 1. Therefore, $V(x) = x^2$ and the drift condition is

$$E[Y^2 \mid x] \le \gamma x^2 + 3/8 \quad \text{for } \gamma \in [1/4, 1).$$

Proposition 2 provides an associated minorization condition on the compact set $C = \{x : x^2 \le w\}$ where $w > 0$.

PROPOSITION 2. *Define probability density $q(\cdot)$ on $\mathbb{R}$ by*

$$q(y) = \varepsilon^{-1} g(y)$$

*where*

$$g(y) = \sqrt{\frac{4}{3\pi}}$$
$$\cdot \exp\left\{-\frac{4}{3}\left(y + \frac{\sqrt{w}}{2}[I(y \ge 0) - I(y < 0)]\right)^2\right\}$$

*and*

$$(6) \qquad \varepsilon = \int g(y)\,dy = 2\Pr\left(Z \le -\sqrt{\frac{2w}{3}}\right)$$

*for $Z \sim N(0,1)$. Then the following minorization condition holds for the transition density $k_x$ of the $x$ chain:*

$$k_x(y) \ge \varepsilon q(y) \quad \text{for all } x \in C$$

*where $C = \{x : x^2 \le w\}$ for $w > 0$.*

PROOF. Let $\pi(\theta \mid x)$ denote the density corresponding to the conditional distribution of $\theta$ given $X = x$. Recall that the transition density $k_x(y)$ can be written as

$$k_x(y) = \int f_\theta(y) \pi(\theta \mid x) \, d\theta$$
$$= \int \sqrt{\frac{2}{\pi}} \exp\{-2(y - \theta)^2\}$$
$$\cdot \sqrt{\frac{4}{\pi}} \exp\{-4(\theta - x/2)^2\}\,d\theta$$

$$= \frac{\sqrt{8}}{\pi} \cdot \sqrt{\frac{\pi}{6}} \exp\left\{-\left(2y^2 + x^2 - 6\left(\frac{x+y}{3}\right)^2\right)\right\}$$

$$\cdot \int \sqrt{\frac{6}{\pi}} \exp\left\{-6\left(\theta - \frac{x+y}{3}\right)^2\right\} d\theta$$

$$= \sqrt{\frac{4}{3\pi}} \exp\left\{-\left(2y^2 + x^2 - 6\left(\frac{x+y}{3}\right)^2\right)\right\}$$

$$= \sqrt{\frac{4}{3\pi}} \exp\left\{-\frac{4}{3}\left(y - \frac{1}{2}x\right)^2\right\}.$$

Also, notice that we can rewrite $C$ as

$$C = \{x : x^2 \leq w\} = \{x : -\sqrt{w} \leq x \leq \sqrt{w}\}.$$

Therefore, for $x \in C$,

$$k_x(y) = \sqrt{\frac{4}{3\pi}} \exp\left\{-\frac{4}{3}\left(y - \frac{1}{2}x\right)^2\right\}$$

$$\geq \sqrt{\frac{4}{3\pi}} \inf_{x \in C} \exp\left\{-\frac{4}{3}\left(y - \frac{1}{2}x\right)^2\right\}$$

$$= \sqrt{\frac{4}{3\pi}} \exp\left\{-\frac{4}{3} \sup_{x \in C}\left(y - \frac{1}{2}x\right)^2\right\}$$

$$= \sqrt{\frac{4}{3\pi}}$$

$$\cdot \exp\left\{-\frac{4}{3}\left(y + \frac{\sqrt{w}}{2}[I(y \geq 0) - I(y < 0)]\right)^2\right\}$$

$$= \varepsilon q(y).$$

It only remains to show that $\varepsilon$ satisfies (6). Let $Z$ represent a standard Normal random variable. Then

$$\varepsilon = \int g(y)\,dy$$

$$= \int_0^\infty \sqrt{\frac{4}{3\pi}} \exp\left\{-\frac{4}{3}\left(y + \frac{\sqrt{w}}{2}\right)^2\right\} dy$$

$$+ \int_{-\infty}^0 \sqrt{\frac{4}{3\pi}} \exp\left\{-\frac{4}{3}\left(y - \frac{\sqrt{w}}{2}\right)^2\right\} dy$$

$$= \Pr\left(\sqrt{\frac{3}{8}} Z - \frac{\sqrt{w}}{2} \geq 0\right) + \Pr\left(\sqrt{\frac{3}{8}} Z + \frac{\sqrt{w}}{2} \leq 0\right)$$

$$= 2\Pr\left(Z \leq -\sqrt{\frac{2w}{3}}\right). \qquad \square$$

Now assume $w > 2L/(1-\gamma) = 3/[4(1-\gamma)]$. Putting Propositions 1 and 2 together immediately guarantees the $x$ chain is geometrically ergodic. These results also allow us to calculate an upper bound on the total variation distance of the $x$ chain to stationarity using Theorem 12 in Rosenthal (1995).

Specifically, for $r = 0.1895820$ (using Rosenthal's notation), $\gamma = 1/4$, and $w = 2.203030$, we have

(7) $\quad \|k_x^l - m\| \leq 0.952697^l + (1.5 + x^2)0.9328785^l$

where $r$, $\gamma$ and $w$ were selected by searching over a grid of candidate values. Now, suppose our goal is to find $n^*$ for which $\|k_0^{n^*} - m\| \leq 0.01$. In this case, Rosenthal's theorem yields $n^* = 99$ since from (7) we have

$$\|k_0^{99} - m\| \leq 0.00980.$$

On the other hand, since $\nu = 0$ and $\sigma^2 + \tau^2 = 1/2$, we can also apply Proposition 4.8 in DKSC to obtain an upper bound on $\|k_0^{n^*} - m\|$. Mainly, Proposition 4.8 gives

(8) $\quad \|k_x^l - m\| \leq \frac{1}{2}\sqrt{\frac{\exp\left(x^2 2^{1-2l}/(1+2^{-2l})\right)}{\sqrt{1-2^{-4l}}} - 1}.$

Thus, when the chain is started at $X_0 = 0$, we find that $n^* = 3$ iterations are sufficient for the $x$ chain to come within 0.01 of the stationary distribution in total variation distance. In fact, using (8) we obtain

$$\|k_0^3 - m\| \leq 0.00552.$$

Clearly, these bounds are sharper than those based on drift and minorization. However, the result based on drift and minorization is still very reasonable. Moreover, it is not clear to us how to apply the techniques of DKSC to the setting where the prior is not conjugate or perhaps is a mixture of conjugate priors. Drift and minorization are flexible enough that they are still applicable in these settings.

## REFERENCES

BAXENDALE, P. H. (2005). Renewal theory and computable convergence rates for geometrically ergodic Markov chains. *Ann. Appl. Probab.* **15** 700–738. MR2114987

FLEGAL, J. M., HARAN, M. and JONES, G. L. (2008). Markov chain Monte Carlo: Can we trust the third significant figure? *Statist. Sci.* To appear.

HOBERT, J. P. and ROBERT, C. P. (2004). A mixture representation of $\pi$ with applications in Markov chain Monte Carlo and perfect sampling. *Ann. Appl. Probab.* **14** 1295–1305. MR2071424

JONES, G. L. (2004). On the Markov chain central limit theorem. *Probab. Surv.* **1** 299–320. MR2068475

JONES, G. L., HARAN, M., CAFFO, B. S. and NEATH, R. (2006). Fixed-width output analysis for Markov chain Monte Carlo. *J. Amer. Statist. Assoc.* **101** 1537–1547. MR2279478

JONES, G. L. and HOBERT, J. P. (2001). Honest exploration of intractable probability distributions via Markov chain Monte Carlo. *Statist. Sci.* **16** 312–334. MR1888447

Jones, G. L. and Hobert, J. P. (2004). Sufficient burn-in for Gibbs samplers for a hierarchical random effects model. *Ann. Statist.* **32** 784–817. MR2060178

Kendall, W. S. (2004). Geometric ergodicity and perfect simulation. *Electron. Commun. Probab.* **9** 140–151. MR2108860

Latuszynski, K. (2008). MCMC ($\varepsilon$-$\alpha$)-approximation under drift condition with application to Gibbs samplers for a hierarchical random effects model. Preprint.

Roberts, G. O. and Rosenthal, J. S. (2004). General state space Markov chains and MCMC algorithms. *Probab. Surv.* **1** 20–71. MR2095565

Roberts, G. O. and Tweedie, R. L. (1999). Bounds on regeneration times and convergence rates for Markov chains. *Stochastic Process. Appl.* **80** 211–229. MR1682243

Rosenthal, J. S. (1995). Minorization conditions and convergence rates for Markov chain Monte Carlo. *J. Amer. Statist. Assoc.* **90** 558–566. MR1340509